\documentclass[twocolumn,floatfix,aps,eqsecnum]{revtex4-2}
\usepackage{graphicx}
\usepackage{amsmath}
\usepackage{amssymb}
\usepackage{bm}
\usepackage{hyperref}

\begin{document}

\title{Topologically protected Casimir effect for lattice fermions}
\author{C. W. J. Beenakker}
\affiliation{Instituut-Lorentz, Universiteit Leiden, P.O. Box 9506, 2300 RA Leiden, The Netherlands}
\date{February 2024}

\begin{abstract}
The electromagnetic Casimir effect has a fermionic counterpart in topological insulators: Zero-point fluctuations of a massless Dirac fermion field mediate a force between magnetic scatterers. The Casimir force is insensitive to disorder that preserves the topological protection of an unpaired Dirac cone. The protection may be broken if the Dirac equation is discretized, and an exponential suppression of the Casimir effect will result if a gap opens at the Dirac point. Here we show how this lattice artefact may be avoided, by applying a recently developed local discretization of the Euclidean action that does not suffer from the fermion-doubling obstruction of local discretizations of the Hamiltonian.
\end{abstract}
\maketitle

\section{Introduction}
\label{intro}

The Casimir effect \cite{Cas48,Plu86,Mos97} is the attractive force between two metal plates in vacuum due to zero-point fluctuations of the electromagnetic field. The radiation pressure is lower in between the plates than outside and pulls the plates together. This generic quantum effect has an electronic analogue for massless Dirac fermions. Early studies of the fermionic Casimir effect were in the context of high-energy physics \cite{Joh75,Amb83,Cas99,Sun04}. The emergence of massless electrons in graphene and topological insulators has created applications in condensed matter physics \cite{Zha08,Shy09,Lu21}.

The Casimir force of a massless field decays as a power law with distance, however, it is suppressed exponentially if the low-energy excitations acquire a nonzero mass, meaning that a gap at the Dirac point opens up in the spectrum \cite{Hay79,Eli03,Nak23}. Dirac fermions remain massless provided that both these conditions are satisfied \cite{Has10,Qi11}:
\begin{enumerate}
\item The low-energy spectrum has only a single Dirac cone (\textit{i.e.}, there is only a single species of low-energy excitations);
\item Chiral symmetry and time-reversal symmetry are not both broken.
\end{enumerate} 
In graphene there are two Dirac cones at opposite corners of the Brillouin zone (valley degeneracy). Hence short-range disorder that couples these two cones via a large momentum transfer can open up a gap at the Dirac point and suppress the Casimir force. In contrast, on the surface of a topological insulator there is a single unpaired Dirac cone. The Casimir effect in a topological insulator can therefore not be suppressed by (time-reversal symmetry preserving) electrostatic disorder. 

This topological protection of the Casimir effect may be compromised if the Dirac equation is discretized \cite{Act00,Ish21,Man22,Flo23}. A local and symmetry-preserving discretization of the Dirac Hamiltonian necessarily introduces a second Dirac cone \cite{Nie81}, an obstruction known as fermion doubling \cite{latticefermions}. One way to work around this obstruction is to embed the Dirac Hamiltonian in a higher dimensional lattice \cite{Kim15,Zie23}. This is how a 3D topological insulator allows for an unpaired Dirac cone on its 2D surface, or how a 2D quantum spin Hall insulator has an unpaired Dirac cone on its 1D edge.

Here we present an alternative, dimension preserving, route to a topologically protected Casimir force for lattice fermions. Following Ref.\ \onlinecite{Zak24} space-time is discretized to obtain an Euclidean action that is local, preserves fundamental symmetries, and has only a single species of low-energy excitations. The second Dirac cone, required by a no-go theorem \cite{Nie81}, is pushed to high energies by the time discretization, and as we will demonstrate does not affect the Casimir force. In particular, we show that short-range disorder has no effect on the power law distance dependence (no exponential suppression because the low-energy excitations remain massless).

The outline of the paper is as follows: In the next section we give the scattering formulation of the fermionic Casimir effect on a space-time lattice, in terms of the tangent discretization of derivative operators \cite{Sta82,Bee23}. The metallic plates of the electromagnetic effect are replaced by ``mass barriers'' \cite{Joh75}, which in a topological insulator correspond to magnetic scatterers. The scattering matrix on the lattice is calculated in Sec.\ \ref{sec_scattering}, and then in Secs.\ \ref{sec_extended}, \ref{sec_spike}, and \ref{sec_2D} the Casimir force is obtained for various cases: extended mass barriers and ``mass spikes'' \cite{Sun04}, in 1D and 2D. The topological protection is demonstrated in Sec.\ \ref{sec_protect}. We conclude in Sec.\ \ref{sec_conclude}.

\section{Casimir effect on a space-time lattice}
\label{freenenergy}

We adapt the scattering formulation of the Casimir effect in the continuum \cite{Jae91,Ken99,Lam06,Ken08,Rah09,Rah11,Mai19} to discrete space and discrete (imaginary) time. We first consider one single spatial dimension (1D case), relevant for the quantum spin-Hall edge, turning later to the 2D case relevant for the surface of a topological insulator.

\subsection{Dirac fermions confined by mass barriers}

We consider the Dirac Hamiltonian
\begin{equation}
{\cal H}=-iv_{\rm F}\sigma_x\partial_x+\mu(x)\sigma_z,
\end{equation}
with $v_{\rm F}$ the Fermi velocity and Pauli spin matrices $\sigma_\alpha$. The Fermi level is fixed at the Dirac point, $E=0$. We set $\hbar$ to unity and denote partial derivatives by $\partial_x\equiv\partial/\partial x$. 

Low-energy excitations are confined to a segment of length $L$ by a pair of mass barriers of length $L_\mu$,
\begin{equation}
\mu(x)=\begin{cases}
\mu_{\rm L}&\text{if}\;\; -L_\mu<x<0,\\
\mu_{\rm R}&\text{if}\;\; L<x<L+L_\mu,\\
0&\text{otherwise}.
\end{cases}\label{mudef}
\end{equation}
On the surface of a topological insulator such a mass profile can be produced by the perpendicular magnetization of a magnetic insulator. The magnetization breaks time reversal symmetry and opens a gap in the spectrum, causing the low-energy excitations to decay for $x<0$ and for $x>L$. In the intermediate region $0<x<L$ the spectrum remains gapless.

The confinement is only effective at energies $E\lesssim \mu_{\rm L},\mu_{\rm R}$. At higher energies the mass barriers are transparent. This is a physical requirement. As we will see shortly, it is also a technical requirement \cite{note3} for our method to work around the fermion doubling obstruction.

The transmission amplitude from one barrier to the other is $t(E)$, the same for transmission from left-to-right and from right-to-left. The reflection amplitudes from the left and right barriers are $r_{\rm L}(E)$ and $r_{\rm R}(E)$, respectively. The product of these scattering amplitudes gives the $L$-dependent contribution to the density of states, according to \cite{Lam06}
\begin{equation}
\begin{split}
&\rho(E)=-\frac{1}{\pi}\operatorname{Im}\frac{d}{dE}\ln[1-\Xi(E+i0^+)],\\
&\Xi(E)=r_{\rm L}(E)r_{\rm R}(E)t(E)^2.
\end{split}\label{rhoE}
\end{equation}

Our objective is to compute the $L$-dependence of the free energy ${\cal F}$, in equilibrium at inverse temperature $\beta=1/k_{\rm B}T$, to obtain the Casimir force $F_{\rm C}=-d{\cal F}/dL$.

\subsection{Tangent fermion discretization}

The free energy has the path integral expression \cite{Mahan,AltlandSimons}
\begin{equation}
\begin{split}
&{\cal F}=-\beta^{-1}\ln Z,\\
&Z=\int {\cal D}\chi\int {\cal D}\bar{\chi}\exp\left(-\int_0^\beta dt\int dx\,{\cal L}[\chi,\bar{\chi}]\right),
\end{split}\label{Fpathintegral}
\end{equation}
in terms of the anticommuting (Grassmann) spinor fields $\chi,\bar{\chi}$ and the Euclidean Lagrangian
\begin{equation}
{\cal L}[\chi,\bar{\chi}]=\bar{\chi}(x,t) (\partial_t+{\cal H})\chi(x,t).
\end{equation}
The Lagrangian is integrated along the interval $0 < it < i\beta$ on the imaginary time axis, with antiperiodic boundary conditions: $\chi(x,\beta) = -\chi(x,0)$. 

In the tangent fermion approach of Ref.\ \cite{Zak24}, space and imaginary time are discretized in units of $a$ and $\tau$, respectively, chosen such that $\beta/\tau$, $L/a$, and $L_\mu/a$ are integer. The space-time lattice consists of the points $it_n = in\tau$, $n = 0,1,2\ldots\beta/\tau -1$, on the imaginary time axis and $x_n = na$, $n\in\mathbb{Z}$ on the real space axis. 

The discretized Lagrangian is given by
\begin{align}
&{\cal L}[\chi,\bar{\chi}]=\bar{\chi}\bigl[-(2/\tau)i\tan(\hat{\omega}\tau/2)+{\cal H}\bigr]\chi,\\
&{\cal H}=(2v_{\rm F}/a)\sigma_x\tan(\hat{k}a/2)+ \mu(x)\sigma_z,\label{Hdef}
\end{align}
with $\hat{k}=-i\partial_x$, $\hat{\omega}=i\partial_t$. The fields $\chi,\bar{\chi}$ are nonlocally coupled by the tangent operators, but a linear transformation produces a local Lagrangian \cite{Zak24}.

The dispersion relation
\begin{equation}
\tan^2(\omega\tau/2)=\gamma^2\tan^2(ka/2),\;\;\gamma=v_{\rm F}\tau/a,
\end{equation}
has two Dirac points in the Brillouin zone: one Dirac point at low energies, $\omega\tau=ka=0$, and a second Dirac point at high energies, $\omega\tau=ka=\pi$. The mass barriers do not confine the Dirac field at high energies, so we can expect that the second Dirac point will not affect the Casimir force. This is how the fermion-doubling obstruction \cite{Nie81} is avoided, without compromising the locality or symmetry of the Lagrangian.

Evaluation of the Gaussian path integral \eqref{Fpathintegral} gives the free energy
\begin{equation}
{\cal F}=-\beta^{-1}\sum_{n=0}^{\beta/\tau-1}\int_{-\infty}^\infty dE\,\rho(E)\ln\bigl[E-(2/\tau)i\tan({\omega_n}\tau/2)\bigr],\label{FfiniteT}
\end{equation}
in terms of a \textit{finite} sum over the Matsubara frequencies $\omega_n = (2n + 1)\pi/\beta$. The pole in the tangent dispersion is avoided by choosing the integer $\beta/\tau$ even.

We substitute Eq.\ \eqref{rhoE} and perform a partial integration,
\begin{equation}
{\cal F}=-\frac{1}{\pi\beta}\operatorname{Im}\sum_{n=0}^{\beta/\tau-1}\int_{-\infty}^\infty dE\,\frac{\ln[1-\Xi(E+i0^+)]}{E-(2/\tau)i\tan(\omega_n\tau/2)}.
\end{equation}
We close the integration interval by a large contour in the upper half of the complex plane, to pick up the poles on the positive imaginary axis. The scattering amplitudes are analytic for $\operatorname{Im}E>0$ (no poles). We thus arrive at
\begin{equation}
\begin{split}
&{\cal F}=-\frac{2}{\beta}\operatorname{Re}\sum_{n=1}^{\beta/2\tau-1}\ln[1-\Xi(i\xi_n)],\\
&\xi_n=(2/\tau)\tan(\omega_n\tau/2).
\end{split}
\end{equation}

In what follows we will limit ourselves to zero temperature, when the sum over the Matsubara frequencies can be replaced by an integral,
\begin{equation}
\lim_{T\rightarrow 0}{\cal F}=-\frac{1}{\pi\tau}\operatorname{Re}\int_0^{\pi}d\omega\,\ln\bigl[1-\Xi\bigl(2i\tan(\omega/2)\bigr].\label{FTzero}
\end{equation}
The continuum formula \cite{Jae91,Ken99,Lam06,Ken08,Rah09,Rah11,Mai19} for the Casimir free energy is obtained if we replace $2\tan(\omega/2)$ by $\omega$ and integrate from $0$ to $\infty$.

\section{Tangent fermion scattering amplitudes}
\label{sec_scattering}

The eigenvalue equation ${\cal H}\Psi=E\Psi$ in the tangent discretization \eqref{Hdef} is nonlocal, it couples the wave function $\Psi_n\equiv\Psi(x=na)$ at arbitrarily distant lattice points. This nonlocality is only apparent \cite{Sta82,Pac21}, it can be removed by the substitution
\begin{equation}
\Psi_n=\tfrac{1}{2}(\Phi_n+\Phi_{n+1}).
\end{equation}
The $x$-dependence of the $\Phi$ field is governed by a \textit{local} relation \cite{Bee23,Two08}, $\Phi_{n+1}=M_n(E)\phi_n$ with transfer matrix
\begin{equation}
\begin{split}
&M_n=\bigl(1-i\sigma_x U_n\bigr)^{-1}\bigl(1+i\sigma_x U_n\bigr),\\
&U_n=(a/2v_{\rm F})\bigl[E-\sigma_z \mu(x=na)\bigr].
\end{split}\label{MnUndef}
\end{equation}

The transfer matrix from $x=0$ to $x=L$ is given by $M_n^{L/a}$ with $\mu\equiv 0$. A right-moving state is an eigenstate of $\sigma_x$ with eigenvalue $+1$, which gives the transmission amplitude
\begin{equation}
t(E)=\left(\frac{1+\tfrac{1}{2}iEa/v_{\rm F}}{1-\tfrac{1}{2}iEa/v_{\rm F}}\right)^{L/a}.\label{tresult}
\end{equation}

The calculation of the reflection amplitude from a mass barrier is a bit more complicated, see App.\ \ref{app_reflection}. For a barrier of length $L_\mu$ and mass $\mu$ we find
\begin{equation}
\begin{split}
&\frac{1}{r(E)}=\frac{E}{\mu}+i\frac{v_{\rm F}\Delta}{a\mu}\;\frac{\left(\frac{2+\Delta}{2-\Delta}\right)^{2L_\mu/a}+1}{\left(\frac{2+\Delta}{2-\Delta}\right)^{2L_\mu/a}-1},\\
&\Delta(E)=\frac{a}{v_{\rm F}}\sqrt{\mu^2-E^2}.
\end{split}\label{rresult}
\end{equation}

The penetration depth $\xi_\mu$ into the barrier at $E=0$ is given by 
\begin{equation}
\frac{a}{\xi_\mu}=\ln\left|\frac{2+\Delta_0}{2-\Delta_0}\right|=\begin{cases}
\Delta_0/4&\text{if}\;\;\Delta_0\ll 1,\\
4/\Delta_0&\text{if}\;\;\Delta_0\gg 1,
\end{cases}\label{ximudef}
\end{equation}
with $\Delta_0=a|\mu|/v_{\rm F}$. The large-$\Delta_0$ behavior is a lattice artefact, only the regime $|\mu|\lesssim v_{\rm F}/a$ is physical.

\section{Casimir force between extended mass barriers}
\label{sec_extended}

In the limit $L_\mu\rightarrow\infty$ of an infinitely extended mass barrier the reflection amplitude \eqref{rresult} simplifies to
\begin{equation}
\lim_{L_\mu\rightarrow\infty}r(i\omega)=i\omega/\mu-i\mu^{-1}\sqrt{\mu^2+\omega^2}.\label{rinfinite}
\end{equation}

\begin{figure}[tb]
\centerline{\includegraphics[width=0.9\linewidth]{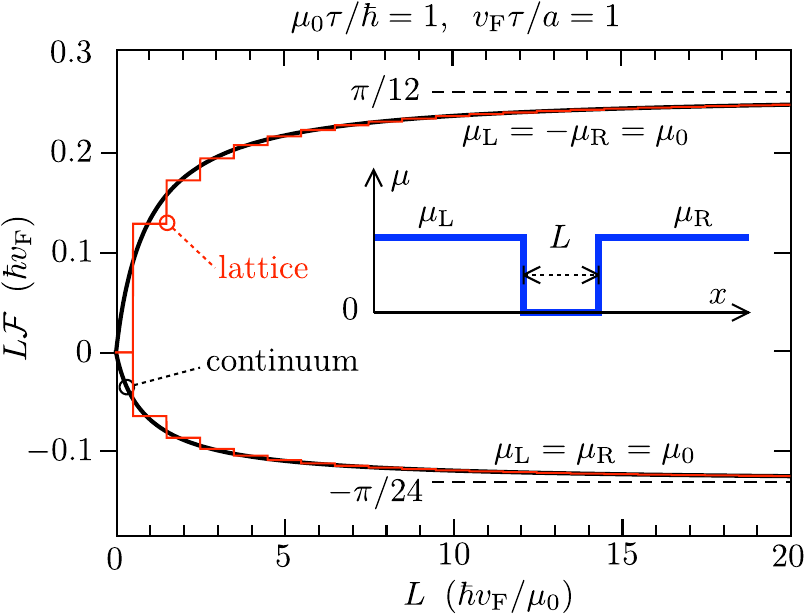}}
\caption{Dependence of the zero-temperature free energy ${\cal F}$ of 1D massless fermions on the separation $L$ of a pair of mass boundaries. The derivative $-d{\cal F}/dL$ is the Casimir force on the barriers. The force is attractive if the two magnetizations $\mu_{\rm L},\mu_{\rm R}$ have the same sign (lower curves), and repulsive if they have the opposite sign (upper curves). The plot compares the result \eqref{FTinfinite} on a lattice (red) with the continuum result \eqref{FTcont} (black). (The dashed lines are the large-$L$ asymptotes \eqref{largeL}.) The space-time lattice constants $a$ and $\tau$ have been chosen such that $\mu_0\tau/\hbar=1=v_{\rm F}\tau/a$. The inset shows the magnetization profile $\mu(x)$ that produces the mass boundaries.
}
\label{fig_barrier}
\end{figure}

We substitute Eqs.\ \eqref{tresult} and \eqref{rinfinite} for the transmission and reflection amplitudes into the free energy formula \eqref{FTzero},
\begin{align}
{\cal F}={}&-\frac{1}{\pi\tau}\int_0^{\pi}d\omega\,\ln\biggl[1+\bigl(\xi/\mu_{\rm L}\tau-\mu_{\rm L}^{-1}\sqrt{\mu_{\rm L}^2+(\xi/\tau)^2}\bigr)\nonumber\\
&\times \bigl(\xi/\mu_{\rm R}\tau-\mu_{\rm R}^{-1}\sqrt{\mu_{\rm R}^2+(\xi/\tau)^2}\bigr)\left(\frac{1-\tfrac{1}{2}\xi/\gamma}{1+\tfrac{1}{2}\xi/\gamma}\right)^{2L/a}\biggr],\label{FTinfinite}
\end{align}
with $\xi=2\tan(\omega/2)$, $\gamma=v_{\rm F}\tau/a$. In Fig.\ \ref{fig_barrier} we compare this with the continuum result,
\begin{align}
{\cal F}_{\rm cont}={}&-\frac{1}{\pi}\int_0^{\infty}d\omega\,\ln\biggl[1+\bigl(\omega/\mu_{\rm L}-\mu_{\rm L}^{-1}\sqrt{\mu_{\rm L}^2+\omega^2}\bigr)\nonumber\\
&\cdot \bigl(\omega/\mu_{\rm R}-\mu_{\rm R}^{-1}\sqrt{\mu_{\rm R}^2+\omega^2}\bigr)e^{-2 \omega L/v_{\rm F}}\biggr].\label{FTcont}
\end{align}
The two match closely.

In the large-$L$ limit Eq.\ \eqref{FTinfinite} tends to
\begin{equation}
{\cal F}_\infty=-\frac{\hbar v_{\rm F}}{\pi L}\int_0^{\infty}dx\,\ln\left(1-r_{\rm L}(0)r_{\rm R}(0) e^{-2 x}\right).\label{Finfinite}
\end{equation}
We thus recover the familiar values \cite{Joh75,Zha08}
\begin{equation}
{\cal F}_\infty=\frac{\hbar v_{\rm F}}{L}\times\begin{cases}
-\pi/24&\text{if}\;\; \operatorname{sign}(\mu_{\rm L}\mu_{\rm R})=+1,\\
+\pi/12&\text{if}\;\; \operatorname{sign}(\mu_{\rm L}\mu_{\rm R})=-1.
\end{cases}\label{largeL}
\end{equation}
The corresponding Casimir force $F_{\rm C}=-d{\cal F}/dL$ decays as $1/L^2$, attractive or repulsive depending on whether $\mu_{\rm L}$ and $\mu_{\rm R}$ have the same or opposite sign.

\section{Casimir force between mass spikes}
\label{sec_spike}

A delta function mass profile, a ``mass spike'' \cite{Sun04}, is represented on the lattice by a mass barrier which is one lattice constant long. This is a model for a magnetic impurity on the quantum spin Hall edge. The reflection amplitude follows from Eq.\ \eqref{rresult} with $L_\mu=a$, which simplifies to
\begin{equation}
r_X(E)=\frac{4 i a\mu_X/v_{\rm F}}{(aE/v_{\rm F}+2 i)^2-(a\mu_X/v_{\rm F})^2},\;\;X\in\{{\rm L,R}\}.\label{rspike}
\end{equation}

\begin{figure}[tb]
\centerline{\includegraphics[width=0.9\linewidth]{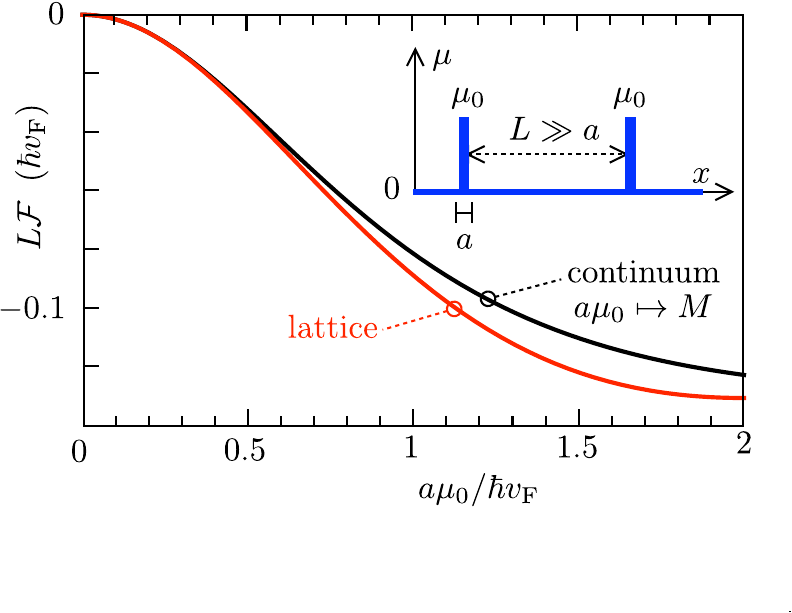}}
\caption{Mass dependence of the Casimir free energy for a delta function mass profile $\mu(x)=M\delta(0)+M\delta(x-L)$ in the continuum [black curve, Eq.\ \eqref{Fspikecont}], modeled by two barriers of height $\mu_0\equiv M/a$ and width $a$  on the lattice [red curve, Eq.\ \eqref{Fspike}]. The agreement is precise for lattice constants $a$ much smaller than both $L$ and $\hbar v_{\rm F}/\mu_0$.
}
\label{fig_spike}
\end{figure}

Substitution into Eq.\ \eqref{Finfinite} gives for $L\gg a$ the Casimir free energy
\begin{equation}
{\cal F}_\infty=\frac{\hbar v_{\rm F}}{2\pi L}\text{Li}_2(-M_{\rm L}M_{\rm R}),\;\;M_{ X}=\frac{4a\mu_{X}/v_{\rm F}}{4+(a\mu_X/v_{\rm F})^2},\label{Fspike}
\end{equation}
with ${\rm Li}_2$ a polylogarithm. This can be compared with the continuum result \cite{Sun04,note1} for the mass profile $\mu(x)=M\delta(x)+M\delta(x-L)$,
\begin{equation}
{\cal F}_{\text{cont}}=\frac{\hbar v_{\rm F}}{2\pi L}\text{Li}_2\bigl(-\tanh^2 (M/v_{\rm F})\bigr).\label{Fspikecont}
\end{equation}
As shown in Fig.\ \ref{fig_spike}, the two expressions agree in the small-mass regime $M\equiv a\mu_0\ll v_{\rm F}$. 

To clarify the correspondence of the lattice and continuum formulas it is helpful to rewrite the lattice formula \eqref{Fspike} for identical mass spikes as 
\begin{equation}
{\cal F}_\infty=\frac{\hbar v_{\rm F}}{2\pi L}\text{Li}_2\bigl(-\tanh^2(a\mu_{\rm eff}/v_{\rm F})\bigr),\;\;\mu_{\rm eff}\equiv v_{\rm F}/\xi_\mu,\label{Fspikexi}
\end{equation}
with $\xi_\mu$ the lattice penetration depth from Eq.\ \eqref{ximudef}. Eq.\ \eqref{Fspikexi} corresponds to the continuum formula \eqref{Fspikecont} if we identify $M/a$ with the effective mass $\mu_{\rm eff}$.

\section{Casimir force between mass barriers on a 2D surface}
\label{sec_2D}

These 1D expressions can readily be generalized to the 2D case. We consider a pair of mass barriers along the $y$-axis, with the mass profile $\mu(x)$ given by Eq.\ \eqref{mudef}. The transmission and reflection coefficients now depend both on energy $E$ and on the transverse wave number $k_y$ (which is a conserved quantity). We work out the case $L_\mu\rightarrow\infty$ of extended barriers.

\begin{figure}[tb]
\centerline{\includegraphics[width=0.9\linewidth]{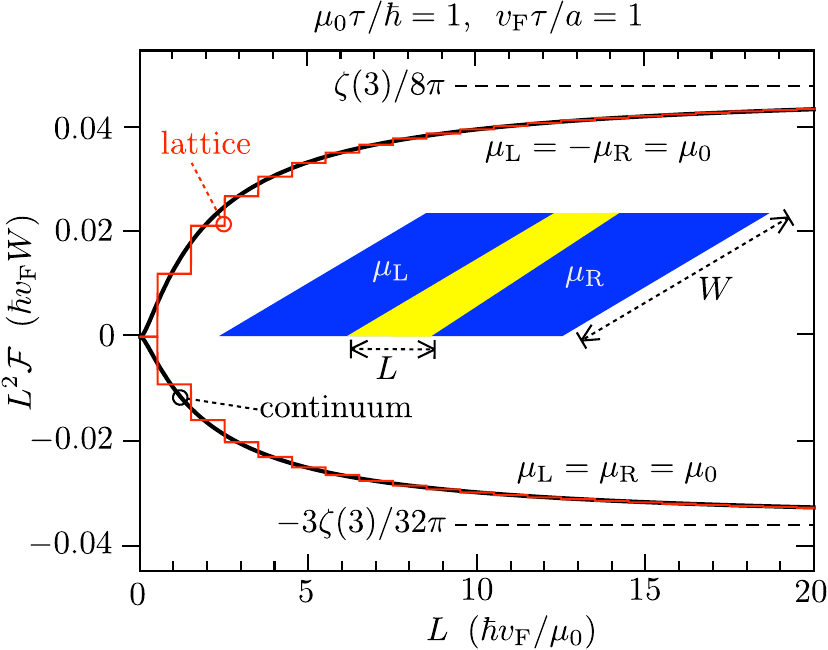}}
\caption{Same as Fig.\ \ref{fig_barrier}, but for the 2D case, computed from Eqs.\ \eqref{calF2D} and \eqref{calF2Dcont}. The dashed lines are the large-$L$ asymptotes \eqref{calF2DlargeL}.
}
\label{fig_barrier2D}
\end{figure}

Using the results for the reflection and transmission coefficients from App.\ \ref{app_reflection} we obtain the free energy
\begin{widetext}
\begin{align}
{\cal F}={}&-\frac{W}{\pi\tau}\int_0^{\pi}d\omega\int_{-\pi/a}^{\pi/a} \frac{dk_y}{2\pi}\,\ln\biggl[1+\frac{1}{\mu_{\rm L}\mu_{\rm R}\tau^2}\biggl(\sqrt{(\gamma {\xi_y})^2+\xi^2}-\sqrt{(\mu_{\rm L}\tau)^2+(\gamma {\xi_y})^2+\xi^2}\biggr)\nonumber\\
&\times \biggl(\sqrt{(\gamma {\xi_y})^2+\xi^2}-\sqrt{(\mu_{\rm R}\tau)^2+(\gamma {\xi_y})^2+\xi^2}\biggr)\left(\frac{2\gamma-\sqrt{(\gamma {\xi_y})^2+\xi^2}}{2\gamma+\sqrt{(\gamma {\xi_y})^2+\xi^2}}\right)^{2L/a}\biggr],\label{calF2D}
\end{align}
with $\xi=2\tan(\omega/2)$ and $\xi_y=2\tan(ak_y/2)$. This expression includes the contributions from both propagating and decaying modes in the inter-barrier region.

In Fig.\ \ref{fig_barrier2D} we compare Eq.\ \eqref{calF2D} with the continuum result,
\begin{align}
{\cal F}_{\rm cont}={}&-\frac{W}{\pi}\int_0^{\infty}d\omega\int_{-\infty}^\infty \frac{d{k_y}}{2\pi} \,\ln\biggl[1+\frac{1}{\mu_{\rm L}\mu_{\rm R}}\biggl(\sqrt{({v_{\rm F}k_y})^2+\omega^2}-\sqrt{\mu_{\rm L}^2+({v_{\rm F}k_y})^2+\omega^2}\biggr)\nonumber\\
&\times \biggl(\sqrt{({v_{\rm F}k_y})^2+\omega^2}-\sqrt{\mu_{\rm R}^2+({v_{\rm F}k_y})^2+\omega^2}\biggr)\exp\left(-\frac{2L}{v_{\rm F}}\sqrt{({v_{\rm F}k_y})^2+\omega^2}\right)\biggr].\label{calF2Dcont}
\end{align}
The two match closely, both tending to the large-$L$ limit of App.\ \ref{app_formulas},
\begin{align}
{\cal F}_\infty={}& -\frac{\hbar v_{\rm F}W}{2\pi L^2}\int_0^{\infty}rdr\,\ln\bigl[1+\operatorname{sign}(\mu_{\rm L}\mu_{\rm R})e^{-2r}\bigr]\nonumber\\
={}&\frac{\hbar v_{\rm F}W}{ L^2}\times\begin{cases}
-3\zeta(3)/32\pi&\text{if}\;\; \operatorname{sign}(\mu_{\rm L}\mu_{\rm R})=+1,\\
+\zeta(3)/8\pi&\text{if}\;\; \operatorname{sign}(\mu_{\rm L}\mu_{\rm R})=-1.
\end{cases}\label{calF2DlargeL}
\end{align}
\end{widetext}

\section{Topological protection of the Casimir force}
\label{sec_protect}

To test the topological protection of the fermionic Casimir effect we apply an electrostatic potential $V(x)$ to the interbarrier region, in the 1D case with extended mass boundaries. This potential preserves time-reversal symmetry, so the fermions should remain massless. 

We consider the staggered potential $V(x)=V_0\cos(\pi x/a)$, so $V=\pm V_0$ on even- and odd-numbered sites, which keeps the Fermi level at zero energy. (A nonzero Fermi wave vector $k_{\rm F}$ would introduce $\sin k_{\rm F} L$ oscillations in the Casimir force \cite{Fuc07,Nak23b}.)

The staggered potential modifies the tangent fermion transmission amplitude by a $\pm V_0$ displacement of the energy,
\begin{align}
t(E)={}&\left(\frac{1+\tfrac{1}{2}i(E-V_0)a/v_{\rm F}}{1-\tfrac{1}{2}i(E-V_0)a/v_{\rm F}}\right)^{L/2a}\nonumber\\
&\times\left(\frac{1+\tfrac{1}{2}i(E+V_0)a/v_{\rm F}}{1-\tfrac{1}{2}i(E+V_0)a/v_{\rm F}}\right)^{L/2a},\label{tresultV}
\end{align}
for $L/a$ an even integer. On the imaginary energy axis we have
\begin{align}
&t\bigl(2i\tan(\omega/2)\bigr)=\nonumber\\
&=\left(1-\frac{8 \sin \omega}{(V_0 a/v_{\rm F})^2 \cos^2( \omega/2)+4 \sin \omega+4}\right)^{L/2a}\nonumber\\
&\rightarrow\exp\left(-\frac{\omega L/a}{1+(V_0 a/2v_{\rm F})^2}\right),
\end{align}
in the large-$L$, small-$\omega$ limit.

The zero-temperature free energy, in the limit $L,L_\mu\rightarrow\infty$, is given by
\begin{equation}
\begin{split}
&{\cal F}_\infty=-\frac{1}{\pi\tau}\int_0^{\infty}d\omega\,\ln\bigl[1+\operatorname{sign}(\mu_{\rm L}\mu_{\rm R})e^{-2\omega L_{\rm eff}/a}\bigr],\\
&L_{\rm eff}=\frac{L}{1+(V_0 a/2v_{\rm F})^2},
\end{split}
\end{equation}
which evaluates to the result \eqref{largeL} with a renormalized length $L\mapsto L_{\rm eff}$.

As anticipated, the power law $L$-dependence of the Casimir force is not affected by the staggered potential in the tangent fermion discretization. In contrast, if a discretization scheme allows a gap $E_{\rm gap}$ to open at the Dirac point, then the Casimir force decays $\propto e^{-E_{\rm gap}L/v_{\rm F}}$ with increasing $L$ \cite{Hay79,Eli03,Nak23}. The gap due to the staggered potential is different for different types of lattice fermions \cite{Bee23,Don22}, of order $V_0$ for naive fermions and of order $V_0^2 a/v_{\rm F}$ for other discretization schemes (Wilson fermions, {\sc slac} fermions, Kogut-Susskind fermions \cite{latticefermions}, see App.\ \ref{app_comparison}).

\section{Conclusion}
\label{sec_conclude}

In summary, we have shown that it is possible to study the fermionic Casimir effect on a lattice without giving up on the topological protection of an unpaired Dirac cone, and without the need to embed the lattice in higher dimensional space. The ingredients that permit to work around the fermion doubling obstruction are two: 1) the tangent fermion space-time discretization that pushes the spurious second Dirac cone to high energies \cite{Zak24}; and 2) the use of the physical condition that high-energy fermions are not confined by mass barriers and therefore do not contribute to the Casimir force.

On the quantum spin Hall edge the Casimir force between magnetic impurities decays slowly $\propto 1/L^2$, unaffected by electrostatic disorder. For a Fermi velocity of $10^6$ m/s this corresponds to an interaction energy of 10~meV at a separation of 10~nm, which may have measurable consequences, such as the aggregation of magnetic impurities with parallel magnetization (as in the analogous case in graphene or carbon nanotubes \cite{Zha08,Shy09}). Our lattice fermion approach should allow for efficient computer simulations of the Casimir effect with electron-electron interactions \cite{Rec05}.

\acknowledgments

This project has received funding from the European Research Council (ERC) under the European Union's Horizon 2020 research and innovation programme.

\appendix

\section{Infinite-mass boundary condition is incompatible with the tangent fermion discretization}
\label{app_infinitemass}

In the infinite-mass limit, when the mass boundaries may be considered impenetrable at all energies, we may replace them by a boundary condition on the modes in the inter-barrier region. That approach is taken in Refs.\ \cite{Ish21,Man22}, for several types of lattice fermions. We have found that the infinite-mass boundary condition is not appropriate for tangent fermions. For that discretization scheme it is essential to retain the physical requirement that the barriers are transparent at high energies. We show this for the 1D case with $\operatorname{sign}(\mu_{\rm L}\mu_{\rm R})=+1$.

The infinite-mass boundary condition for tangent fermions implies the energy quantization
\begin{equation}
\begin{split}
&E_m=(2/\tau)\tan\bigl((m+1/2)\pi a/L\bigr),\\
&m=-L/a,\ldots ,-1,0,1,\ldots L/a-1.
\end{split}
\end{equation}
The zero-temperature free energy \eqref{FfiniteT} is then given by
\begin{widetext}
\begin{align}
{\cal F}={}&-\frac{1}{2\pi\tau}\sum_{m=-L/a}^{L/a-1}\int_0^{2\pi}d\omega\,\ln\bigl[2\gamma\tan\bigl((m+1/2)\pi a/2L\bigr)-2i\tan({\omega}/2)\bigr]\nonumber\\
={}&-\frac{2L}{a\tau}\ln 2+\sum_{m=0}^{L/a-1}f(m+1/2),\;\;f(x)=-\frac{2}{\tau}\ln\big( 1+\gamma\tan\bigl(x\pi a/2L\bigr)\bigr).\label{Ftangentmass}
\end{align}
The regularized zero-point energy is the difference between sum and integral,
\begin{equation}
\delta {\cal F}=\sum_{m=0}^{L/a-1}f(m+1/2)-\int_0^{L/a}f(x)\,dx={\cal F}-L\lim_{L\rightarrow\infty}L^{-1}{\cal F}.
\end{equation}

We use the Abel-Plana formula \cite{note2}
\begin{subequations}
\begin{align}
&\sum_{n=\lceil A-\nu\rceil}^{\lfloor B-\nu\rfloor}f(n+\nu)-\int_{A}^{B}f(x)\,dx=\Delta(A,\nu)+\Delta(B,\nu)-Q(A,\nu)+Q(B,\nu),\\
&\Delta(A,\nu)=\begin{cases}
\tfrac{1}{2}f(A)&\text{if}\;\;A-\nu\in\mathbb{Z},\\
0&\text{otherwise},
\end{cases}\\
&Q(A,\nu)=\frac{1}{i}\int_0^\infty dy\,\left[\frac{f(A+iy)}{e^{2\pi y-2\pi i(A-\nu)}-1}-\frac{f(A-iy)}{e^{2\pi y+2\pi i(A-\nu)}-1}\right],
\end{align}\label{AbelPlana}
\end{subequations}
for $A,B,\nu\in\mathbb{R}$.

Applied to Eq.\ \eqref{Ftangentmass} (with $A=0$, $B=L/a$, $\nu=1/2$) this gives the zero-point energy
\begin{align}
\delta {\cal F}&=-\frac{4}{\tau}\int_0^\infty dy\,\frac{\arctan[\gamma\tanh(y\pi a/2L)]}{e^{2\pi y}+1}+\frac{4}{\tau}\int_0^\infty dy\,\frac{\arctan[\gamma\operatorname{cotanh}(y\pi a/2L)]}{e^{2\pi y}+1}\nonumber\\
&=\frac{1}{\tau}\ln 2-\frac{v_{\rm F}}{L}\frac{\pi  \left( \gamma^2+1\right)}{24 \gamma^2}+{\cal O}(L^{-2}).
\end{align}
\end{widetext}
The $L$-independent offset does not contribute to the Casimir force and can be ignored. The term that decays as $1/L$ only attains the correct prefactor \eqref{largeL} when $\gamma=v_{\rm F}\tau/a\rightarrow\infty$, so for vanishing lattice constant $a$. For finite $\gamma$ the tangent fermion discretization overestimates the Casimir force in the infinite-mass limit, because of spurious contributions from high-energy modes at the edge of the Brillouin zone. This complication is avoided in the scattering approach from the main text, because there the mass barriers are transparent at high energies, hence the spurious modes are not confined and do not contribute to the Casimir force.

\section{Scattering of tangent fermions by a mass barrier}
\label{app_reflection}

We compute the tangent fermion reflection and transmission matrices for a mass barrier. We consider a 2D system, such as the surface of a topological insulator, with the barrier oriented along the $y$-axis. The mass profile $\mu(x)$ depends only on $x$, so momentum $k_y$ parallel to the boundary is conserved.

The transfer matrix $M_n$ from $x=na$ to $x=(n+1)a$ is given by \cite{Bee23,Two08}
\begin{subequations}
\begin{align}
M_n={}&\left(1-i\sigma_x U_n-\tfrac{1}{2}\sigma_z \xi_y\right)^{-1}\left(1+i\sigma_x U_n+\tfrac{1}{2}\sigma_z \xi_y\right),\\
U_n={}&(a/2v_{\rm F})\bigl[E-\sigma_z \mu(x=na)\bigr],\;\;\xi_y=2\tan(ak_y/2).
\end{align}
\end{subequations}
(The 1D Eq.\ \eqref{MnUndef} corresponds to $k_y= 0$.) The full transfer matrix through the barrier, of length $L_\mu$ an integer multiple of the lattice constant $a$, is $M_n^{L_\mu/a}$.

The eigenstates of $M_n$ for $\mu=0$ are
\begin{equation}
\begin{split}
&\chi_+=2^{-1/2}\left(\sqrt{1-(\gamma\xi_y/E\tau)^2}-i\gamma\xi_y/E\tau,1\right),\\
&\chi_-=2^{-1/2}\left(1,-\sqrt{1-(\gamma\xi_y/E\tau)^2}+i\gamma\xi_y/E\tau\right).
\end{split}
\end{equation}
For $|\gamma\xi_y/E\tau|<1$ these are states which carry the same current in opposite directions, 
\begin{equation}
\langle\chi_\pm|\sigma_x|\chi_\pm\rangle=\pm\sqrt{1-(\gamma\xi_y/E\tau)^2},
\end{equation}
so they can serve as the basis of incoming and outgoing states for the scattering matrix.

We transform the transfer matrix to the new basis,
\begin{equation}
\tilde{M}_n=\Omega^{-1}M_n\Omega,\;\;\Omega=\begin{pmatrix}
\chi_+\\
\chi_-
\end{pmatrix}^\top,
\end{equation}
and then find the scattering matrix $S$ from
 \begin{align}
&(\tilde{M}_n)^{L_\mu/a}=\begin{pmatrix}
m_{11}&m_{12}\\
m_{21}&m_{22}
\end{pmatrix}\Rightarrow\\
& S=\begin{pmatrix}
-m_{21}/m_{22}&1/m_{22}\\
m_{11}-m_{12}m_{21}/m_{22}&m_{12}/m_{22}
\end{pmatrix}\equiv \begin{pmatrix}
r&t\\
t'&r'
\end{pmatrix}.\nonumber
\end{align}

After some algebra we thus obtain the reflection and transmission coefficients
\begin{align}
&\frac{1}{r}=\frac{1}{r'}=\frac{\varepsilon}{\mu}+i\frac{v_{\rm F}\Delta}{a\mu}\frac{{Q}^2+1}{{Q}^2-1},\label{rapp}\\
&\frac{1}{t}=\frac{1}{t'}=\tfrac{1}{2}({Q}+1/{Q})-\tfrac{1}{2}i\frac{a\varepsilon}{v_{\rm F}\Delta}({Q}-1/{Q}),\label{tapp}
\end{align}
with the definitions
\begin{subequations}
\begin{align}
&\varepsilon=E\sqrt{1-(\gamma\xi_y/E\tau)^2},\\
&\Delta=\frac{a}{v_{\rm F}}\sqrt{\mu^2+(\gamma\xi_y/\tau)^2-E^2},\\
&{Q}=\left(\frac{2+\Delta}{2-\Delta}\right)^{L_\mu/a}.
\end{align}
\end{subequations}
For $k_y=0$ Eq.\ \eqref{rapp} reduces to Eq.\ \eqref{rresult}.

We note the large-$L_\mu$ limit of the reflection coefficient for $E=i\omega$ on the imaginary axis,
\begin{align}
&\lim_{L_\mu\rightarrow\infty}r(i\omega)=\lim_{{Q}^2\rightarrow\infty} r(i\omega)\nonumber\\
&\;=\frac{i\omega}{\mu}\sqrt{1+(\gamma\xi_y/\omega\tau)^2}-\frac{i}{\mu}\sqrt{\mu^2+(\gamma\xi_y/\tau)^2+\omega^2},\label{rinfiniteapp}
\end{align}
generalizing Eq.\ \eqref{rinfinite} to $k_y\neq 0$.

The transmission coefficient through a massless region of length $L$ follows from Eq.\ \eqref{tapp} with $\mu=0$,
\begin{equation}
\lim_{\mu\rightarrow 0}t(E)=\left(\frac{1+\tfrac{1}{2}i\varepsilon a/v_{\rm F}}{1-\tfrac{1}{2}i\varepsilon a/v_{\rm F}}\right)^{L/a},
\end{equation}
which for $k_y=0$ reduces to Eq.\ \eqref{tresult}.

\section{Fermionic Casimir force on widely separated mass boundaries}
\label{app_formulas}

To compare with the literature on the fermionic Casimir force we record the limiting expressions we obtain for the case $L_\mu,L\rightarrow\infty$ of extended and widely separated mass barriers. 

In the scattering formulation the zero-point energy in $d+1$-dimensional continuous space-time is given by an integral over the imaginary frequency $i\omega$ and a $d-1$-dimensional integral over the transverse wave vector $\bm{k}_\parallel$,
\begin{align}
&{\cal F}_\infty=-\hbar\int_0^\infty\frac{d\omega}{\pi}\int\frac{d\bm{k}_\parallel}{(2\pi)^{d-1}}\nonumber\\
&\quad\times\ln\left[1-r_{\rm L}(i\omega,\bm{k}_\parallel)r_{\rm R}(i\omega,\bm{k}_\parallel)e^{-2L\sqrt{(\omega/v_{\rm F})^2+|\bm{k}_\parallel|^2}}\right].
\end{align}
This is the zero-temperature free energy per unit area of the mass barriers and for a single spin degree of freedom. 

In the large-$L$ limit we may replace the reflection coefficients $r_{\rm L},r_{\rm R}$ by their value $\pm i$ at $\omega=0=\bm{k}_\parallel$, where the sign $\pm i$ depends on the sign of the mass. Upon transformation to spherical coordinates we have the integral
\begin{align}
{\cal F}_{\infty}={}&-\frac{\hbar v_{\rm F} }{L^{d}}\frac{2\pi^{d/2}}{(2\pi)^d\Gamma (d/2)}\int_0^\infty r^{d-1}dr\nonumber\\
&\times\ln\left[1+\operatorname{sign}(\mu_{\rm L}\mu_{\rm R})e^{-2r}\right],
\end{align}
which evaluates to
\begin{align}
{\cal F}_\infty={}&\frac{\hbar v_{\rm F}}{L^d}\frac{  \zeta(d+1) \Gamma (d)}{2^{3 d-1}\pi^{d/2}\Gamma (d/2)}\nonumber\\
&\times\begin{cases}
(1-2^{d})&\text{if}\;\; \operatorname{sign}(\mu_{\rm L}\mu_{\rm R})=+1,\\
2^d&\text{if}\;\; \operatorname{sign}(\mu_{\rm L}\mu_{\rm R})=-1.
\end{cases}\label{Finftyd}
\end{align}

Specifically, the coefficient $c_d^\pm$ in ${\cal F}_\infty=c_d^\pm\hbar v_{\rm F}/L^d$ for $d=1,2,3$ and $\operatorname{sign}(\mu_{\rm L}\mu_{\rm R})=\pm 1$ equals
\begin{equation}
\begin{split}
&\{c_1^+,c_2^+,c_3^+\}=\biggl\{-\frac{\pi }{24},-\frac{3 \zeta (3)}{32 \pi },-\frac{7 \pi ^2}{5760}\biggr\},\\
&\{c_1^-,c_2^-,c_3^-\}=\biggl\{+\frac{\pi }{12},+\frac{\zeta (3)}{8 \pi },+\frac{\pi ^2}{720}\biggr\}.
\end{split}
\end{equation}
The Casimir force $F_{\rm C}=-d{\cal F}/dL$ on the barriers is attractive if $\operatorname{sign}(\mu_{\rm L}\mu_{\rm R})=+1$ and repulsive if $\operatorname{sign}(\mu_{\rm L}\mu_{\rm R})=- 1$.

The formula \eqref{Finftyd} for the attractive fermionic Casimir force was derived in Ref.\ \onlinecite{Cas99} by a different method (zeta function regularization). The result for the repulsive case equals $-2$ times the bosonic result from Ref.\ \onlinecite{Amb83}.

\section{Comparison with other lattice fermions}
\label{app_comparison}

As derived in Ref.\ \onlinecite{Don22}, a staggered potential $V(x)=V_0\cos(\pi x/a)$ may cause a gap $E_{\rm gap}$ to open up at the zero-energy Dirac point. This causes an exponential suppression $\propto e^{-E_{\rm gap}L/v_{\rm F}}$ of the Casimir force \cite{Hay79,Eli03,Nak23}. For tangent fermions the Dirac fermions remain massless. We summarize the gap results for various other types \cite{latticefermions} of 1D lattice fermions.

The staggered potential couples states at $k_x$ and $k_x+\pi/a$, as described by the Hamiltonian
\begin{equation}
H_{V}(k_x)=\begin{pmatrix}
H(k_x)&V_0/2\\
V_0/2&H(k_x+\pi/a)
\end{pmatrix}.
\end{equation}
For a given lattice Hamiltonian $H$ one thus obtains the following gap $E_{\rm gap}$ at $k_x=0$ in the spectrum of $H_V$:
\begin{widetext}
\begin{itemize}
\item{\em naive fermions:} ${H}(k_x)=( v_{\rm F}/a)\sigma_x\sin ak_x\Rightarrow E_{\rm gap}=V_0$. 
\item{\em Wilson fermions:} ${H}(k_x)=( v_{\rm F}/a)\sigma_x\sin ak_x+m_0\sigma_z(1-\cos ak_x)\Rightarrow E_{\rm gap}=\sqrt{4 m_0^2+V_0^2}-2 m_0$.
\item{\em Kogut-Susskind fermions:} ${H}(k_x)=( v_{\rm F}/a)[\sigma_x\sin ak_x+\sigma_y(1-\cos ak_x)]\Rightarrow E_{\rm gap}=\sqrt{4 v_{\rm F}^2/a^2+V_0^2}-2 v_{\rm F}/a$.
\item{\em SLAC fermions:} ${H}(k_x)=-i( v_{\rm F}/a)\sigma_x \,\ln e^{iak_x}\Rightarrow E_{\rm gap}=\sqrt{\pi^2 v_{\rm F}^2/a^2+V_0^2}-\pi v_{\rm F}/a$.
\end{itemize}
The resulting Casimir force decays as $e^{-L/\xi}$ with $\xi\propto 1/V_0$ for naive fermions and $\xi\propto 1/V_0^2$ in the other three cases.

All of this should be contrasted with
\begin{itemize}
\item{\em tangent fermions:}
${H}(k_x)=2(v_{\rm F}/a)\sigma_x\tan (ak_x/2)\Rightarrow E_{\rm gap}=0$,
\end{itemize}
where the Casimir force retains the power law decay $\propto 1/L^2$.
\end{widetext}


\begin{thebibliography}{99}
\bibitem{note3} As an alternative to the scattering approach to the Casimir effect, one may assume infinitely high barriers and replace the reflection amplitude by a boundary condition on the confined fields. As we discuss in App.\ \ref{app_infinitemass}, that alternative approach is not compatible with the tangent fermion discretization.
\bibitem{note1} Eq.\ \eqref{Fspikecont} follows upon integration of Eq.\ 4.19 from Ref.\ \onlinecite{Sun04}, in the massless fermion limit $m\rightarrow 0$. Comparison with the earlier Eq.\ 4.17 indicates a typo in Eq.\ 4.19, the factor $(t^2-m^2)$ in the denominator of the integrand should be replaced by $(t^2-m^2)\sinh^2\lambda$, so that the integrand becomes $-\frac{2 td \sinh^2\lambda}{\pi e^{2  td} \cosh ^2\lambda +\pi  \sinh ^2\lambda}$ in the $m\rightarrow 0$ limit.
\end{thebibliography}
\end{document}